\documentclass[twoside]{article}
\usepackage[a4paper]{geometry}
\usepackage[latin1]{inputenc} % ou \usepackage[utf8]{inputenc}
\usepackage[T1]{fontenc} % ou \usepackage[OT1]{fontenc}
\usepackage{RR}
\usepackage{hyperref}
\RRNo{}
\RRdate{September 2012}
\RRauthor{% les auteurs
Emmanuel Nataf\thanks{Universit{\'e} de Lorraine}%
  \and
Olivier Festor%\thanks{Footnote for second author}%
}
\authorhead{Nataf \& Festor}
\RRtitle{{\'E}stimation en ligne de la Dur{\'e}e de Vie des Batteries pour les R{\'e}seaux de Capteurs sans Fil}
\RRetitle{Online Estimation of Battery Lifetime for Wireless Sensor Network}
\titlehead{Online Battery lifetime}
\RRabstract{Battery is  a major hardware component  of wireless sensor
  networks.  Most of  them  have  no power  supply  and are  generally
  deployed  for a  long time.  Researches  have been  done on  battery
  physical  model and  their adaptation  for sensors.   We  present an
  implementation   on  a   real  sensor   operating  system   and  how
  architectural constraints  have been assumed.  Experiments have been
  made  in  order  to  test  the  impact of  some  parameter,  as  the
  application throughput, on the battery lifetime.}

\RRkeyword{battery, wireless sensors network, Contiki, Cooja}
\RRresume{La  batterie  est  un  composant matériel  majeur  dans  les
  réseaux de capteurs sans fil.  Des recherches ont été menées sur les
  modèles  physique   des  batteries  et  leur   adaptation  pour  les
  capteurs. Nous présentons une  implantation sur un véritable système
  d'exploitation de capteurs et comment les contraintes d'architecture
  ont pu être pris en  compte.  Des expériences ont ensuite été menées
  pour  tester  l'influence  de  certains paramètres  comme  le  débit
  d'application sur la durée de vie des batteries}

\RRmotcle{batterie, réseaux de capteurs sans fil, Contiki, Cooja}
\RRprojets{Madynes}
 \URLorraine % pour ceux qui sont \`a l'est
\begin{document}
\bibliographystyle{plain}
\makeRR   % cas d'un rapport de recherche
%% \makeRT % cas d'un rapport technique.
%% a partir d'ici, chacun fait comme il le souhaite
\section{Introduction}

Wireless  Sensors Networks  have  a growing  place  in industrial  and
domestical domains. Generally sensors need to be small, cheap and easy
to deploy on any  physical environment. These constrained networks are
made  possible by the  processor size  reduction and  high performance
battery.  Nevertheless these networks have  to be care of their energy
consumption  because they  are usually  deployed  for a  long time  or
because  sensors  are  so  small  that their  battery  has  a  limited
capacity. The growing interest  on battery lifetime estimation for WSN
appears    in   the    standardization   work    on    these   network
\cite{rfc6550}. They defines a  routing protocol with one metric being
the remaining battery level \cite{rfc6551}.

Battery  models  that relate  its  use  with  its lifetime  have  been
proposed   since  embedded  systems   (laptop,  cellular   phone)  are
widespread.  All of them take into account two phenomenons occuring in
battery cell;  the {\it  Rate Capacity Effect}  and the  {\it Recovery
  Effect}  \cite{panigrahi01}.  The first  gives the  energy consumed
under a constant current load,  when transmitting for example, and the
second  gives the energy  recovered during  inactivity or  low current
load.  At a  constant current  load,  oxidation at  the anode  induces
reduction at the cathode. The reduction decreases the concentration of
positive ions near the cathode and so the available energy. But during
inactivity or low  current load the positive ions  near the anode have
time to move toward the  cathode, thus increasing the available energy
and  so  the  battery  lifetime.   A  stochastic  model  is  given  in
\cite{panigrahi01},  but the more  accurate modeling  is given  by the
analytical  equation  (\ref{rakmod})  \cite{rakhmatov01,  rakhmatov02,
  rao03}.  $\alpha$ denotes the  total battery capacity that is equals
to the  current load $i(\tau)$ consumed  since the use  of the battery
until the time  $L$ (the battery $L$ifetime) and  the current that was
unavailable because  positive ion concentration was  not sufficient at
the cathode.   The $\beta$ parameter  is the diffusion  coefficient of
electroactive species inside the battery.
\begin{equation}
\alpha = \int_0^Li(\tau)d\tau + 2\sum_{m=1}^{\infty}\int_0^Li(\tau)e^{-\beta^2m^2(L-\tau)}d\tau
\label{rakmod}
\end{equation}

Use  of this model  can be  found in  \cite{rakhmatov03,seunking05} to
schedule    tasks    for   embedded    system    or   to    simulation
\cite{Timmermann03}.  It is widely used for example by a circuit-based
battery   model   \cite{zhang10}   or   in  WSN   routing   simulation
\cite{sausen10}.  The model can not  be implemented as is on WSN nodes
albeit  some  results  achieved  using  the  first  part  of  equation
(\ref{rakmod})  \cite{dunkels07, kerasiotis10}. In  so doing,  they do
not take  into account  the recovery process  that occurs  during idle
time despite it is generally up to $90\%$ of the WSN node timelife.

The contribution of \cite{rahme10} is an approximation of the equation
(\ref{rakmod})  that  can be  implemented  on  WSN  node.  Its  global
equation (\ref{recmod}) recursively computes $\sigma(L_n)$ that is the
consumed  charge in  {\sl  mA-min} at  time  $L_n$.  This  computation
depends on  the consumed charge  at time $L_{n-1}$, the  time interval
$\Delta=L_n - L_{n-1}$ being constant.
\begin{equation}
\sigma(L_n) = \sum_{k=1}^{n}I_k \delta_k + \lambda\left(\sigma(L_{n-1}) - \sum_{k=1}^{n-1}I_k \delta_k\right)\\ 
+ 2I_{n}A(L_n, L_{n-1} + \delta_k, L_{n-1})
\label{recmod}
\end{equation}
With this approach,  one can know the remaining  charge in the battery
at time $L_n$ by the  difference between $\sigma(L_n)$ and the initial
battery  parameter  $\alpha$  of  the  equation  (\ref{rakmod}).   The
recovery effect of the battery is computed through the function $A$ of
the equation (\ref{recmod})  that is approximated by the  use of an $f$
function:
\begin{equation}
A(L_n, L_{n-1} + \delta_k, L_{n-1}) = \\
\sum_{m=1}^{\infty}\frac{e^{-\beta^2m^2({\bf L_n-L_{n-1} - \delta_k})}-e^{-\beta^2m^2(L_n- L_{n-1})}}{\beta^2m^2} \\
\simeq \frac{f(\nu)}{\beta^2} - \frac{f(\Delta)}{\beta^2}
%\label{A1}
\end{equation}
This approximation  depends on  the idle time  of the mote  during the
$\Delta$ interval, given  by the difference $\nu =  \Delta - \delta_k$
($\nu$ is  the idle  time of the  $L_{n-1}$ period and  $\delta_k$ its
activity  time).  Following  \cite{rahme10}, the  $f$ function  can be
also  approximated  by  the   use  of  the  equations  (\ref{fnu})  to
(\ref{c0}) (see \cite{rahmerr, rahme09} for details):
\vspace{5pt}
\noindent
\begin{equation}
\frac{f(\nu)}{\beta^2}     =      \frac{\nu}{2}     -     \frac{\sqrt
  \pi}{\beta}.\sqrt{\nu} + \frac{\pi^2}{6 \beta^2}
\label{fnu}
\end{equation}

\begin{equation}
\sqrt{\nu} \simeq \sqrt{a} + \frac{\nu - a}{2\sqrt{a}}
\label{rnu}
\end{equation}

\begin{equation}
\frac{f(\Delta)}{\beta^2} = \sum_{m=1}^{\infty}\frac{e^{-\beta^2m^2\Delta}}{\beta^2m^2} = c_0
\label{c0}
\end{equation}

The  $\beta$   parameter  (physical  diffusion   coefficient)  in  the
equations (\ref{fnu}) and (\ref{c0})  is a constant value that depends
on the battery and is computed from several tests of discharge and the
least   squares  estimation   method  (along   with   $\alpha$).   The
approximation of  the equation (\ref{rnu}) concerns  $\sqrt{\nu}$ by a
mean value : $a$, computed offline.  In our case the mean value $a$ is
fixed at $90\%$  of $\Delta$, that is the value  we measure on several
tests.  The last approximation  of the equation (\ref{c0}) is computed
with  the $m$  parameter  ranging from  $1$  to $10$  because the  sum
quickly converges.

The  recursivity  of the  model  depends  on  the $\lambda$  parameter
defined    as   the   ratio    $\frac{A(L_{n+1},   \delta_1,0)}{A(L_n,
  \delta_1,0)}$  for each $n$  but that  can be  bounded by  the value
$e^{-\beta^2\Delta}$  computed off line.

Our contribution  in this paper  is to fill  the last gap  between the
approximated model and its implementation within an existing operating
system  for sensors.   We use  physical nodes  to validate  our energy
consumption  retrieval  process and  observe  in  simulated world  how
battery reacts  face to  network events and  configuration parameters.
The remainder of this paper  is organized as follow.  The next section
is  focalized on  the  elementary  operations of  WSN  node and  their
current draw.  Details of the  implementation are given in the section
\ref{impsec} and  first results are shown  in the last  part.  We give
some conclusions and plan future works at the end of the document.

\section {Current draw}
\label{model}

\subsection{Linear draw}

 The  first term  in equation  (\ref{recmod}) is  the sum  of products
 between  current load $I_k$  and time  interval $\delta_k$  since the
 begining of the  battery livetime (when $k=1$) until  the time $L_n$.
 The $\delta_k$  time interval  is a sub  interval of  $\Delta$ during
 which the  battery was used by  the node ($\delta_k  \le \Delta$). To
 compute this value we have to  know which components of the motes use
 the current along the time and at which current rate.

Mote's current  information can  be obtained from  datasheet document.
For example,  the table  \ref{curtab} shows the  current load  for two
mote types  (Sky or WSN430  motherboard, respectively with  CC2420 and
CC1100 communication chips).  The columns  are the usual states of the
duty  cycle for  any  mote :  CPU,  LPM, TX  and  RX respectively  for
processor, low  power mode, transmitting and  receiving.  Currents are
given  in milli-Ampere ($mA$).   The given  values for  TX and  RX are
linked to signal power and throughput configuration of the mote.
\begin {table}[htbp]
\begin{center}
\begin{tabular}{|c||c|c|c|c|}\hline
Mote & CPU & LPM & TX & RX \\ \hline \hline
Sky & $1.8$ & $0.0545$ & $17.4$ & $18.8$\\ \hline
Wsn430 & $2$ & $0.02$ & $16.1$ & $15.2$\\ \hline
\end{tabular}
\end{center}
\caption{Motes current draw}
\label{curtab}
\end{table}

\begin{figure*}[htbp]
\begin{center}
\includegraphics[scale=.58]{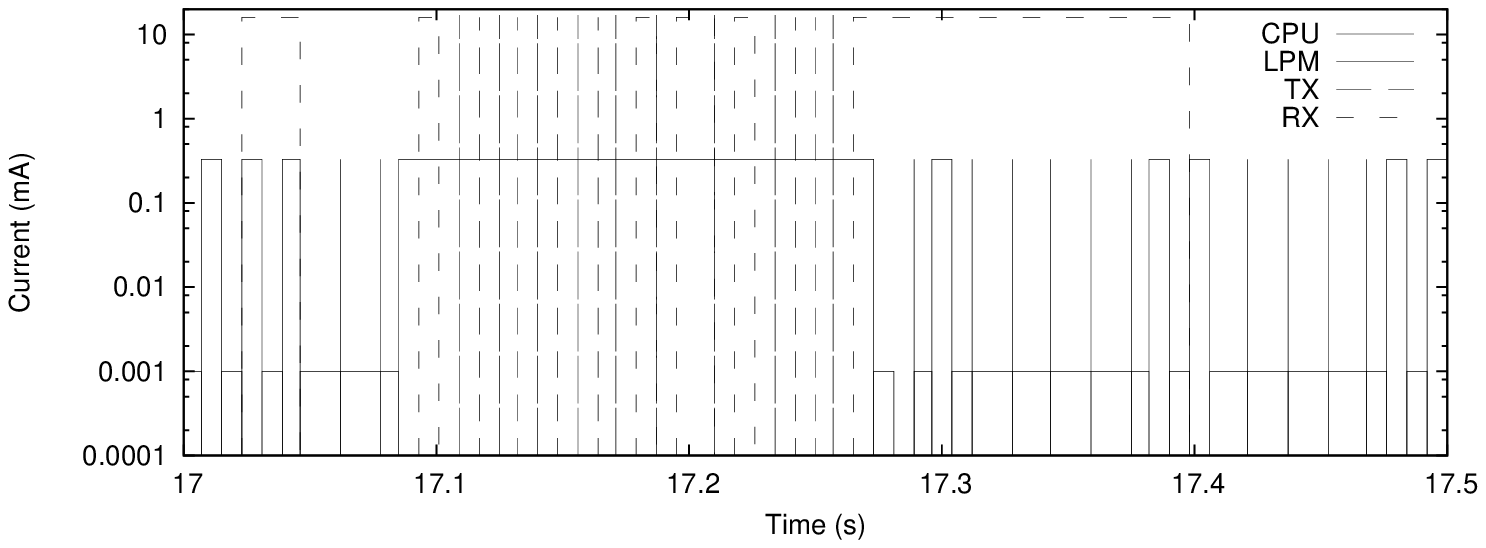}
\includegraphics[scale=.58]{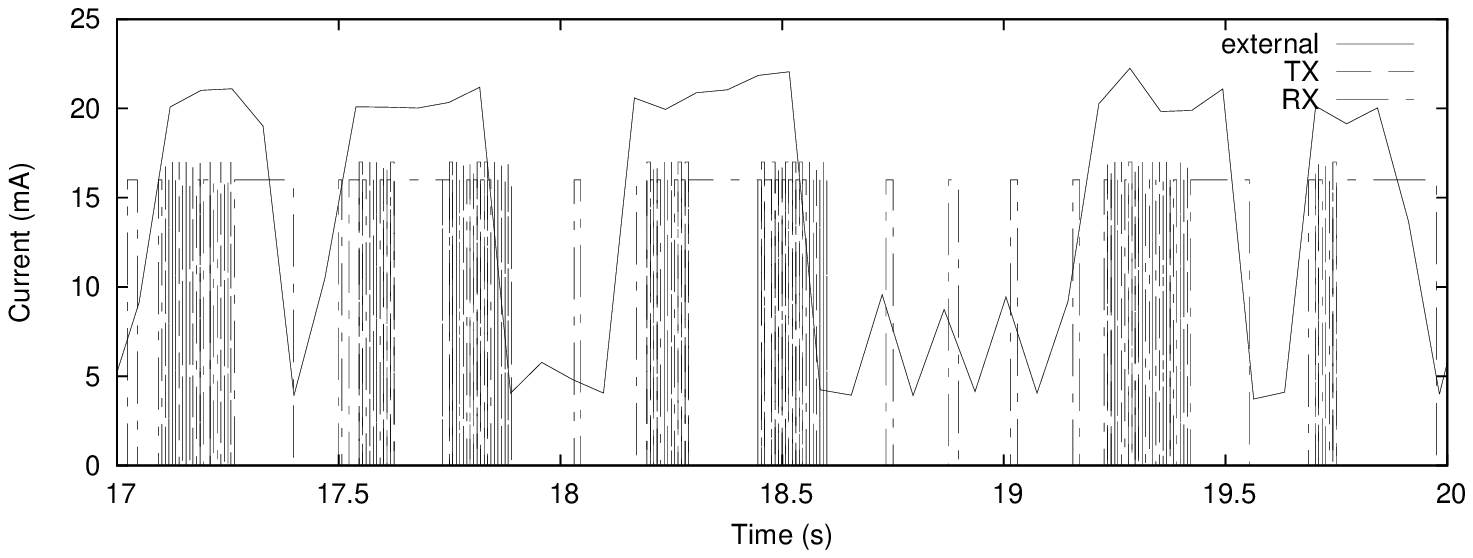}
\end{center}
\caption{Mote modes and currents}
\label{duties}
\end{figure*}

The left part of figure \ref{duties}  shows a short time interval of a
real WSN430 mote with the Contiki operating system \cite{contiki04} on
the senslab testbed \cite{senslab11}.  We plot these states by logging
start  and stop  time  on mote  serial  output that  is redirected  by
senslab to a TCP connexion.   These times are accessibles by an energy
related application programming interface on Contiki.  The CPU and LPM
are not overlapping themselves and  fill all the time duration.  These
states are related  to the mainboard activities and  so either the CPU
is  active or the  low power  mode is  raised.  TX  and RX  states are
related to the  radio activities and they are  neither used every time
nor they overlap  each otther.  One can note that  TX mode only occurs
during CPU and that RX mode begins during CPU but can spread on LPM.

The senslab testbed  also provides an external polling  of the current
consumed by  a mote  during an experiment.   The right part  of figure
\ref{duties}  shows  the  curves  of  measured  current  at  the  mote
endpoints and the duties cycles as  in the left part (we just leave TX
and RX mode that have the main current values). One can see that start
and stop times of duties match well with measured current.

Consequently,   the   current  draw   of   a  $\Delta=\delta_{CPU}   +
\delta_{LPM}$  intervall is  the result  of the  equation (\ref{Ikdk})
where each $C_{state}$ is the current load (cf.table \ref{curtab}) and
each  $\delta_{state}$ is the  sum of  all state  sub-intervals inside
$\Delta$.
\begin{equation}
I_k\delta_k = C_{CPU}.\delta_{CPU} + C_{LPM}.\delta_{LPM} + C_{TX}.\delta_{TX} + C_{RX}.\delta_{RX}
\label{Ikdk}
\end{equation}

\section{Implementation constraints}
\label{impsec}

Even  with  the  help  of  \cite{rahme10},  we  must  be  care  of  mote
limitations  on  code size  and  arithmetics  capabilities. The  strong
constraint is  that MSP430  can not handle  floating point  values but
only  signed or unsigned  integer.  

\subsection{Offline computation}

Approximations given make use of $\sqrt{\pi}$ or $\pi^2$ and we had to
round them with a new unit  as it is show on table \ref{conv}. Most
of them  are just a factor  of thousand to remove  floating point with
saving  enougth  precision.  The  computed  values  $\nu$,  $c_0$  and
$\lambda$ are computed  with a $\Delta$ of $2s$.

\begin{table}[htbp]
\begin{center}
\begin{tabular}{|l|l|l|l|l|l|l|l|l|} \hline
&$\pi^2$ & $\sqrt{\pi}$ & $\beta$ & $c_0$ & $\lambda$ & $a$ & $\sqrt{a}$ & $\frac{1}{2\sqrt{a}}$\\ \hline \hline
R & 9.869 & 1.772 & 1 & 1.337 & 0.967 & 0.03 & 0.173 & 2.886\\ \hline
I & 9869 & 1772 & 10  & 1337 & 967 & 30 & 173 & 2886\\ \hline
\end{tabular}
\end{center}
\caption{Real to Integer values}
\label{conv}
\end{table}

\subsection{Online computation}

We  compute the  consumed current  every $\Delta$  time  interval.  The
values $\delta_{CPU, LPM, TX, RX}$ are provided in milli second and we
use  the current  load  of the  table  \ref{curtab} with  a factor  of
thousand.  The result  is in  $10^{-3}mA.ms$  and it  is converted  in
$mA.mn$.

 We  compute $\nu$ with  the low power
mode period and the radio use :
\begin{equation}
\nu = \left\{
  \begin{array}{l l}
    \delta_{LPM} - (\delta_{TX}  + \delta_{RX}) & \quad \textrm{if  $\delta_{LPM} > (\delta_{TX}  + \delta_{RX})$}\\
    0& \quad \textrm{else}\\
  \end{array} \right.
\end{equation}
If the radio is  more used than the CPU is idle  then there is no idle
time for the battery.

The  varying part  of the  recovery  function $A$  (equation \ref{fnu})  is
rewritten in order to lose as little precision as possible :
\begin{equation}
\frac{f(\nu)}{\beta^2}   =   \frac{10^4\beta^2\nu   +  \pi^22.10^3   -
  12\beta\sqrt{\pi}\sqrt{\nu}}{\beta^212.10^5}
\end{equation}
At this step we  had to use a $64$ bits intermediary  value to get the
result (MSP430  compiler allows such  data type but they  are actually
emulated with several $32$ bits values).

Finaly, the remaining energy is computed with the values above and the
preceding remaining energy value  (a $\Delta$ before).  We compute the
ratio with $255$ as $100\%$. This value is given in RPL routing metric
recommandation \cite{rfc6551} and we use the present work to implement
and test  an energy-based  routing plane in  an other work  beyond the
scope of  this paper. However we  compute the remaining  energy with a
precision of $5$ numbers, that  is from $255.10^5$, so one can observe
energy consumption at a very fine coarse.

The code size  of the presented implementation is  about $12Kb$ on the
$50Kb$  allowed by our  processor.  At  the memory  level, we  save on
memory six $32$  bits values from one computation  to the other (times
of  CPU,  LPM,  TX,  RX  and  previous  values  for  $\sigma(L_{n-1})$
and $\sum_{k=1}^{n-1}I_k\delta_k$).

\section{Simultation Results}

For our simulation  we use the Cooja WSN  simulator \cite{cooja06} and
test a network  of nodes build on the sky  mote platform.  These nodes
use  IPv6 and the  RPL routing  protocol to  perform a  simple collect
application  during  which  one  or  more nodes,  the  {\it  Senders},
periodically send sensing informations  to a receiving node called the
{\it Sink},  configured to be  the root of  the RPL routing  tree. All
node have an  initial battery charge of $880mAh$.   {\it Senders} send
one data packet  every second and use the  {\it contikiMac} radio duty
cycle  \cite{ctkmac11}.  We use  a  linux  {\it  Ubuntu 11.0}  desktop
computer with a $3 GHz$ {\it Intel} processor and $4 Gb$ of memory.

\subsection{Booting node}

\begin{figure*}[htbp]
\begin{center}
\begin{tabular}{cc}
\includegraphics[scale=.56]{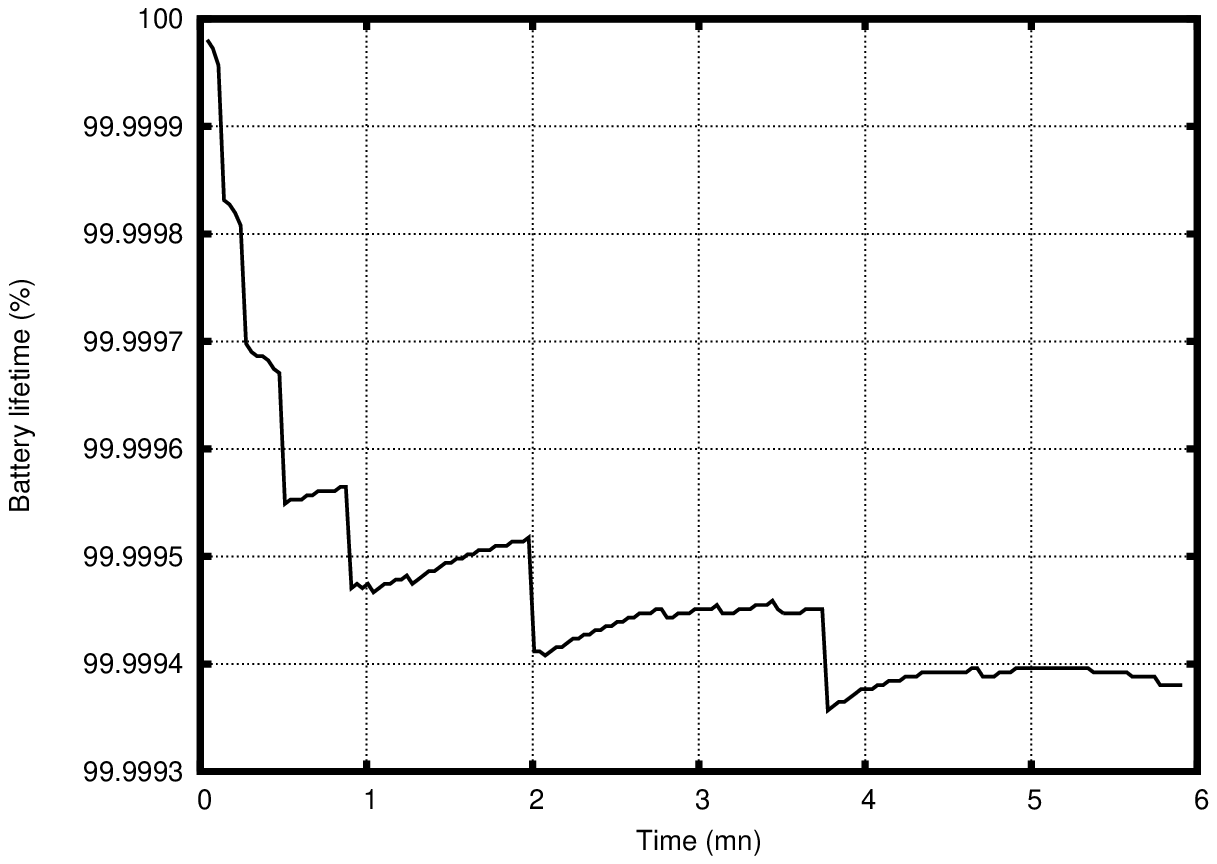}&\includegraphics[scale=.56]{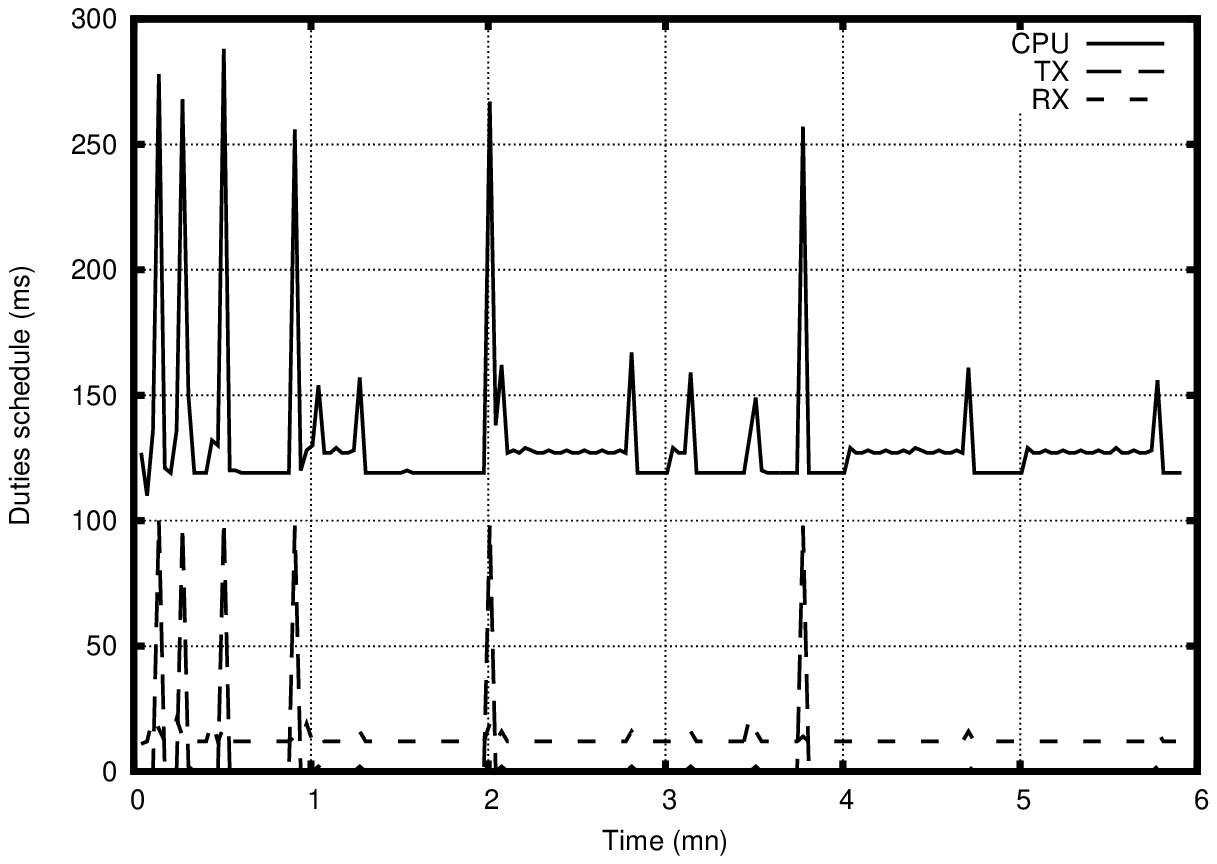}\\
(a)&(b)
\end{tabular}
\caption{First minutes of battery lifetime}
\label{init}
\end{center}
\end{figure*}

We  start our  experiment with  the  observation of  the node  booting
process.   The figure \ref{init}  puts together  the evolution  of the
battery lifetime and the time spends by activities of the node. During
the first  minute, on the part (a),  one can note a  large decrease of
the  battery\footnote{that  is  only   $0.0005$\%  of  the  total}  to
correlate  with the  strong  activity on  the  part (b)  for the  same
period.   Mainly  this  activity  is  related to  system  and  network
initialization (remains  the strong power for  TX and RX  of the table
\ref{curtab}).
The recovery  of battery is visible  at this very  close level, during
the second minute  and the two following ones.  These recovery periods
match with less activities period.

\subsection{Network building}

\begin{figure*}[htbp]
\begin{center}
\begin{tabular}{cc}
\includegraphics[scale=.56]{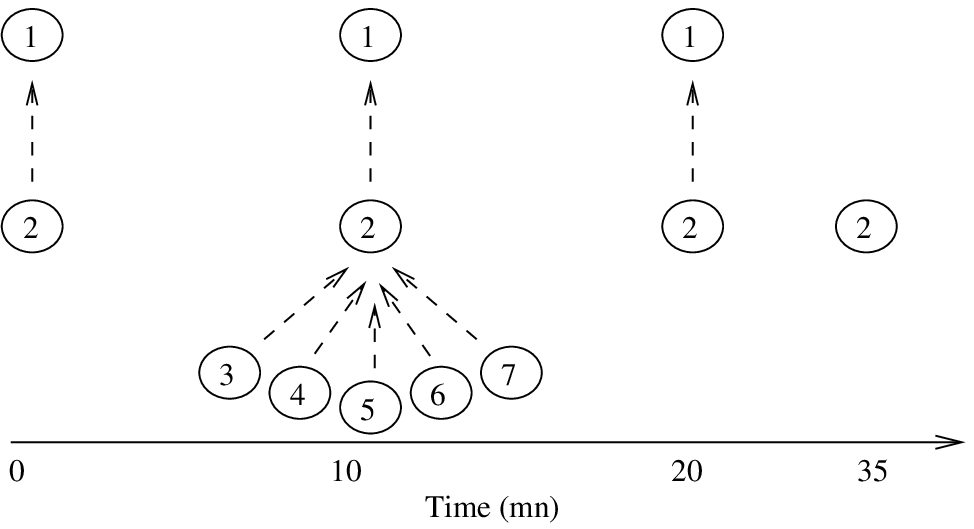}&
\includegraphics[scale=.56]{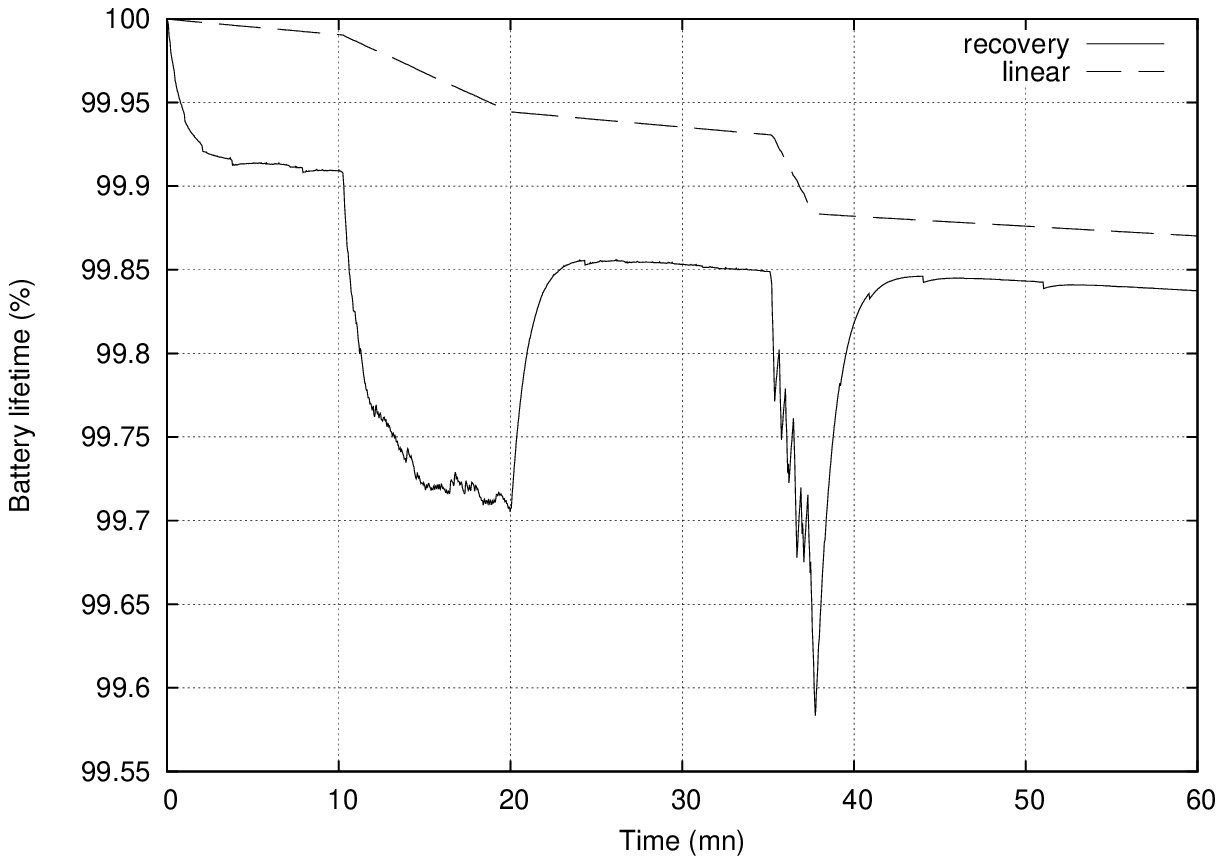}\\
(a)&(b)
\end{tabular}
\caption{Battery lifetime and RPL}
\label{scenari}
\end{center}
\end{figure*}

We  continue the  simulation  following the  figure \ref{scenari}  (a)
where after ten minutes of simulated time, we add five nodes ``under''
the non-root node to observe the cost of the routing tree building and
traffic cumulation. The  node $1$ is the {\it Sink}  and all other are
{\it Sender}.  We  leave these nodes sending and  forwarding (only for
the  node $2$) data  packets during  ten minutes  and remove  the five
nodes previously added. This can  occurs in real life for example with
mobile nodes or with perturbed communication environment. A last point
of  interest, at  the thirty  fifth minute,  is when  a node  lost its
destination node (its default next hop router) because the network has
to build itself again.

The figure  \ref{scenari}(b) plots  the remaining battery  lifetime of
the node  $2$ along  our experiment. The  {\tt recovery} curve  is our
implementation and the  {\tt linear} is a simple  additive function of
current draw. The first strong  decrease of the tenth minute is mainly
due to  traffic overhead  and very few  is from RPL  control messages.
Once the routing plane is established, between the tenth and twentieth
minute, the battery decreases with a greater slope and with more micro
variations.  When removing nodes at the twentieth minute, the node $2$
shows a quick and strong recovery of its battery lifetime. As given by
the model, the fall of current  draw lets the battery recovers some of
its  current charge. During  the fiftheen following minutes  the node
sends one packet per second to the {\it Sink} and the curve is similar
to  the begining  (without booting  process)  but at  a lower  battery
level.

At the thirty  fifth minute we remove the parent of  the node $2$. The
very large decrease of battery is again mainly related to a strong use
of communication. The  node has no more IPv6  routeur neighbor then it
polls the  network with neighbour  discovery messages and  after three
minutes, the node decides to build a new network and just send few RPL
control message.  One can there observe a very significant recovery at
the part (b).

The  linear  model   is  very  less  reactive  and   never  grows,  as
expected.  This  strong difference  suggest  it  could  over or  under
estimate the lifetime depending on how the other sensors behave.

\subsection{Implementation Accuracy}

\begin{figure}[htpb]
\includegraphics[scale=.65]{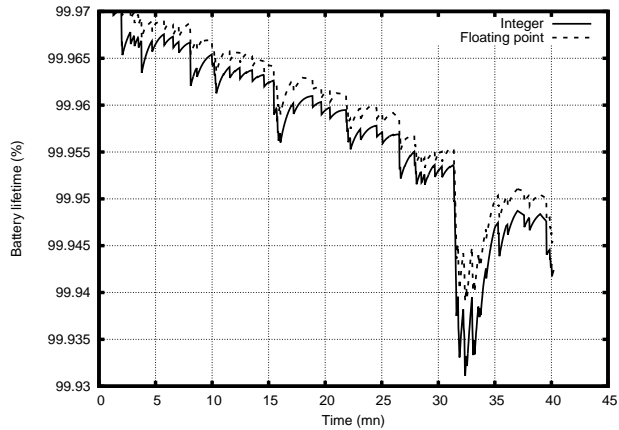}
\caption{Floating point vs Integer battery lifetime computation}
\label{rounding}
\end{figure}

The approximation we  made on basic values and  the rounding effect of
divide  operations lead  to  residual errors  all  along the  lifetime
estimation. We have implemented a  floating point version of the model
with a perl script and compare its result with the computation made by
the mote.   The figure \ref{rounding} shows  a snapshot of  a WSN like
those  of the  figure \ref{scenari}.a  with the  two  battery lifetime
estimations. The mote computes values  a bit lower than floating point
but  the difference  stays  constant. Strong  variations (between  the
minutes $30$ and  $35$) have a much more amplitude  but do not deviate
the lifetime.
These  results enforce  our  confidence on  our  implementation as  it
follows the theory of  battery consumption and especially the recovery
one. We continue research on the simulation for long time duration.

\subsection{Network life}
\label{netlifesec}

Estimation  of sensor  lifetime  is a  prior  information for  their
deployement.   Sensors networks  can  be in  place  for several  years
before be  replaced or recharged and  so battery should  be chosen to
match the  expected duration.

\begin{figure*}[htbp]
\begin{center}
\begin{tabular}{cc}
\includegraphics[scale=.56]{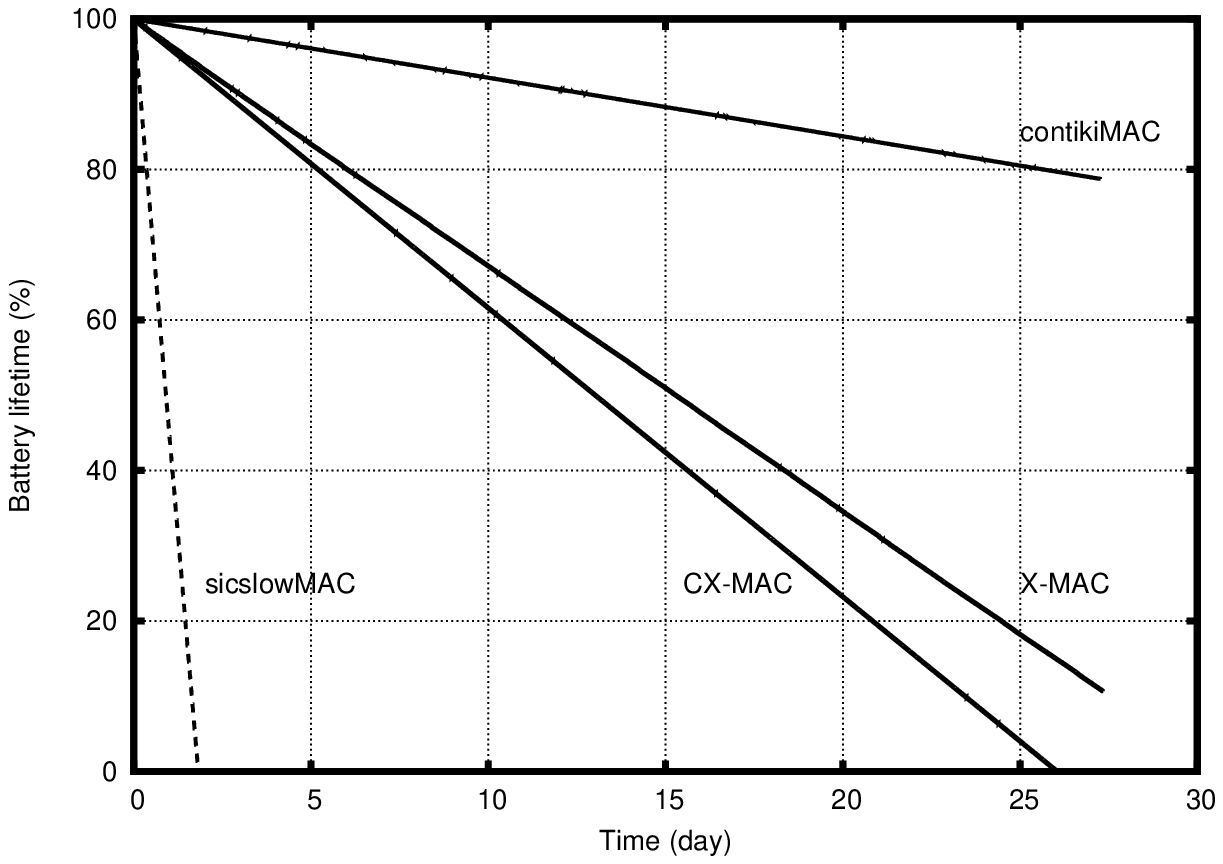}&\includegraphics[scale=.56]{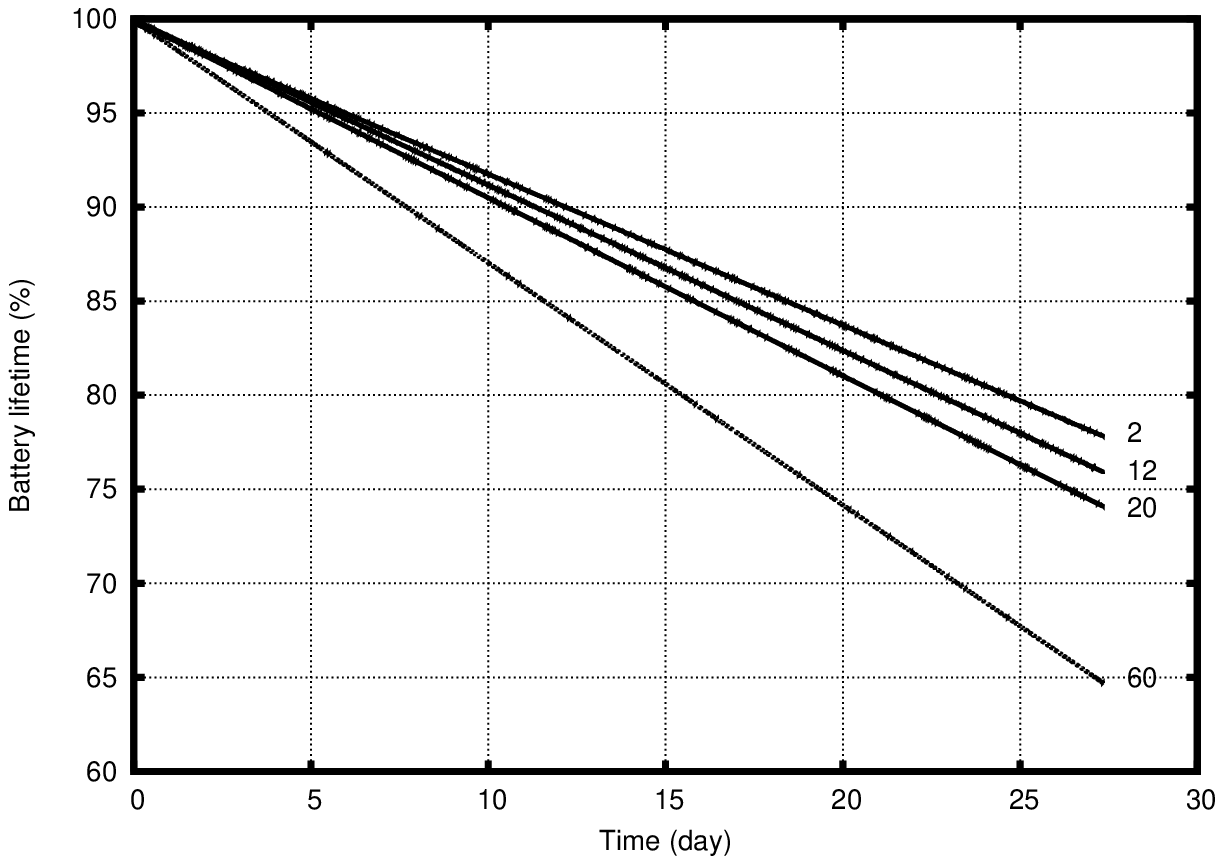}\\
(a)&(b)\\
\end{tabular}
\caption{Network life}
\label{netlife}
\end{center}
\end{figure*}

The  Contiki operating  system provides  some Radio  Duty  Cycle (RDC)
implementations  we  have  used  in  order  to  compare  their  energy
consumption.   The part (a)  of the  figure \ref{netlife}  shows these
experiments. All these RDC are asynchronous and packet oriented.
\begin{itemize}
\item  the  contikiMAC  \cite{ctkmac11}  allows the  greatest  battery
  lifetime. It can sleep up to  $99\%$ of the time and was measured as
  been ten times less energy  consumer than the following X-MAC.  When
  a node has to send, it transmits the packet several times, until the
  receiver return an acknowledge.
\item  X-MAC \cite{buettner06} is  a more  consumer; the  sender sends
  several  preambule  packets until  acknowledge  reception and  then
  sends the data packet.
\item CX-MAC  (Compatibility X-MAC)  is a variation  of X-MAC  that is
  provided by Contiki to be usable on more radio chips.
\item sicslowmac  (also in Contiki) handles $802.15.4$  frames and has
  its radio alway  open. Note that with  a TX at $\simeq 20  mA$ and a
  battery  at $880 mAh$,  the theory  gives $\frac{880}{20}=44  h$ (cf
  table \ref{reglin}).
\end{itemize}

We have  also measured the  throughput effect on battery  lifetime and
plot them in part (b)  of the figure \ref{netlife}. The contikiMAC RDC
was used because it is the  less battery consumer and so the impact of
the throughputs is more discernible.   Numbers within the part (b) are
the  throughput  of  the  application,  from $60$  packets  by  minute
($pkts/mn$) to  $1$.  Each  packet has  a size of  $87$ bytes  and the
application  is  parameterised by  a  number  of  seconds between  two
packets.  The figure  shows that the lifetime is  roughsly linear with
the throughput  but other tests  with less throughput than  $1 pkt/mn$
have no  significantly improved the  lifetime. The network  traffic is
then  negligeable against  sensor  base activity  (indeed the  routing
protocol is the main radio user).

 These figures are  very close to a line (only a  zoom shows their are
 not) and we use a  linear regression with the least square estimation
 technique to  envision the time  when the battery will  be completely
 empty (i.e.  with $0\%$  of remaining energy). The table \ref{reglin}
 contains  the battery lifetime  estimation for  each lines  in figure
 \ref{netlife}.  The  short lifetimes of these networks  (no more than
 four months) is related to the light battery we have used. Battery of
 node like Sky  are usually around $2000 mAh$ and  should allow WSN be
 operational for  one year. Nevertheless real batteries  have a secure
 cut off level  that prevent them to be  fully discharged because they
 will be damaged and not able to be charged again.

\begin{table}[htbp]
\begin{center}
\begin{tabular}{|c|c|}\hline
RDC & Battery lifetime  \\
&(days)  \\\hline\hline
contikiMAC&  128\\ \hline
X-MAC & 30 \\ \hline
CX-MAC & 26\\ \hline
sicslowMAC & 1.8\\ \hline \hline
\multicolumn{2}{|c|}{Throughput : $1 pkt/mn$} \\ \hline
\end{tabular}
\begin{tabular}{|c|c|}\hline
Throughput & Battery lifetime  \\
($pkts/mn$) & (days) \\ \hline\hline
60 & 77 \\ \hline
20 & 105\\ \hline
12 & 113\\ \hline
2 & 124\\ \hline \hline
\multicolumn{2}{|c|}{RDC : contikiMAC} \\ \hline
\end{tabular}
\caption{Battery life estimation}
\label{reglin}
\end{center}
\end{table}

\section{Conclusions}

This  paper presents  a  first implementation  of  a battery  lifetime
estimation inside  an existing  operating system for  WSN. Based  on a
recognized  theoritical battery model,  our work  has show  how sensor
internal architecture impacts on concrete realization. Our simulations
help us to better understand how sensor use their battery in bootstrap
process or during networking. We believe this work is usefull for many
applications  as routing  optimization  that is  a  work we  currently
plan. We also  have to launch long time  tests establishing a relation
between  the   reamaining  energy   and  the  cut   off  level   of  a
battery. Moreover  several battery technologies  have to be  tested to
enforce our implementation model.

\bibliography{bcharge}

% that's all folks
\end{document}